\documentclass[aps,prl,twocolumn,preprintnumbers,showpacs]{revtex4}
\usepackage{graphicx}
\usepackage{amsfonts}
\usepackage{amsmath}
\usepackage{amssymb}
\usepackage{amsfonts}

\providecommand{\mb}[1]{\mathbf{#1}}
\providecommand{\mc}[1]{\mathcal{#1}}

\providecommand{\norm}[1]{\lvert#1\rvert}
\newcommand{\ud}{\mathrm{d}}

\begin{document}
\bibliographystyle{unsrt}
\title[Injection]{A new theoretical approach to 1:1 electrolytes
at low temperature}
\author{Weimin Zhou$^1$ and Jerome K. Percus$^{1,2}$}
\affiliation{ $^{1} $Department of Physics, New
York University, 4 Washington Place, New York, New York 10003\\
$^{2}$Courant Institute of Mathematical Sciences, New York
University, New York, NY, 10012}
\begin{abstract}
A new theoretical approach to 1:1 electrolytes at low temperature
is developed, RPM and SAPM are studied with this approach, and
their critical points of first order phase transition are
calculated.
\end{abstract}

\volumeyear{year} \volumenumber{number} \issuenumber{number}
\eid{identifier}
\date[Date:]{\today}
\received[Received text:]{\today}

\maketitle This report presents a new theoretical approach to
ionic systems at low temperature. The systems under study are
primitive models of electrolytes: Equal numbers, $N$, of
positively and negatively charged hard spheres of diameter $a_+$
and $a_-$ in a volume $V$, carrying charges $+q_0$ and $-q_0$
respectively, interact via the Coulomb potential $\pm{}q_0^2/Dr$
in a medium of dielectric constant $D$. The size asymmetry of $+$
and $-$ ions is measured by $\lambda=a_-/a_+$, with $\lambda=1$
being the special case of the Restricted Primitive model (RPM) and
otherwise the Size Asymmetric Primitive Model (SAPM). Such systems
undergo a first order phase transition at low temperature and
moderately low density, they have received increasing attention in
recent years (see, e.g. \cite{Stell, Fisher, Enrique, SAPM,
Enrique2}). In the following, we define $a=\frac{1}{2}(a_++a_-)$
and the normalized reciprocal temperature $\beta={q_0^2}/(k_BTD)$,
which has a length dimension, we denote the total particle number
density by $\rho$, and because of the symmetry with respect to the
exchange of $+$ and $-$ ions, only $\lambda\le{}1$ is considered.

\begin{figure}[t]
\includegraphics[width=0.5\textwidth]{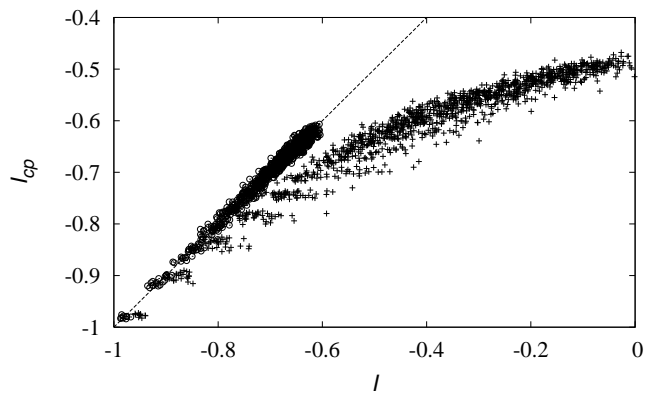}
\caption{$I_{cp}$ vs. $I$ for 1000 random sampling configurations
at $\rho=0.05$. cross: with no restrictions on the selection of
configurations; circle: with DNR imposed on the selection of
configurations; dashed line: $I$, as reference.\label{CPTestFig}}
\end{figure}

There are $2N^2-N$ pairwise interactions in the system, among
which $N^2$ are negative (between unlike ions), $N^2-N$ are
positive (between like ions). Intuitively, if the $N^2-N$ positive
interactions could be used to cancel $N^2-N$ negative
interactions, we would only have $N$ negative pairwise
interactions left in the system. And this is desirable because
it's much more manageable to deal with $N$ than $2N^2-N$
interactions, analytically or numerically. In this context, let's
review a pairing procedure first introduced by Stillinger and
Lovett \cite{Stillinger}. In their procedure, all the distances
between two unlike ions are computed, the first pair is defined as
the two unlike ions that have the closest distance, and this step
is repeated, taking into account only ions that remain unpaired,
until all the ions in the system are exhausted. If a certain
prescription\cite{MyThesis} is used on the situations (of zero
weight) when one ion has more than one unlike ion at the same
distance, the result of such procedure is that each positive ion
has one and only one negative ion as its partner, and vice versa.
We call this procedure "Closest Pairing" (CP). The following
numerical study was carried out to exam the energy profile of CP
pairs: $N=500$, $a_+=a_-=1$, ions are randomly \cite{MyThesis}
placed in a cubic box of volume $V$, with hard wall boundary
condition; pairs are formed through CP procedure, and the $N$
attractive inter-ion interactions of CP pairs are summed up,
divided by $N$, denoted by $I_{cp}$, and then compared with $I$:
the sum of all the $2N^2-N$ pairwise interactions divided by $N$.
We found that for a wide range of densities, at least for
$\rho\in[0.01,0.2]$, $I_{cp}$ is within close range of $I$ for
configurations with $I\in[-1,-\rho^{1/3}]$; while for
$I>-\rho^{1/3}$, $I_{cp}$ has a value less but around
$-\rho^{1/3}$ and shows no significant change for different
configurations; and of course, $I_{cp}$ cannot have values below
$(-1/a)$, here, -1. A typical plot of $I_{cp}$ vs. $I$ is shown in
Figure~\ref{CPTestFig}.

It is then adequate to approximate $I$ by $I_{cp}$ in the low
temperature regime: $\beta\gg{}a$, when, expectedly, most unlike
ions are closely paired up and the value of $I$ is in the close
neighborhood of $(-1/a)$. For clarity of presentation, we call
such approximation "CP Approximation (CPA)". But configurations
with $I>-\rho^{1/3}$ do pose a potential problem for us. Although
in the thermodynamic limit when $N, V \to\infty$, such
configurations would have virtually no contribution to any
physical phenomenon we are interested here, they do occupy a
relatively very large volume in phase space because they allow
particles to move much more freely. Thus if no precautions are
taken in the calculation of the partition function with $I$
approximated by $I_{cp}$, this large but ought to be insignificant
phase space volume would be associated with a significantly
magnified Boltzmann factor, and this could significantly affect
the result quantitatively, if not qualitatively. In order to
prevent this from happening, we put a restriction on the selection
of the allowed configurations, called "Detailed Negativeness
Requirement" (DNR): \textit{two like ions can not get closer than
a distance $\mathfrak{D}$ unless at least one of them has at least
one unlike ion at a position that is closer than $\mathfrak{D}$.}
A similar numerical study was carried out in which only the
configurations that comply with DNR are sampled, and we found that
DNR does eliminate unwanted configurations \cite{MyThesis}; a
corresponding plot of $I_{cp}$ vs. $I$ is also presented in
Figure~\ref{CPTestFig} (we see that DNR also limits local energy
fluctuations and admittedly deserves further study).

In the following, we present an analytical scheme that employs CPA
with DNR.

In the Grand Canonical Ensemble, the system is described by five
parameters $(a,\lambda,z_+,z_-,\beta)$. The partition function is:
\begin{equation}\label{Partition}
\begin{split}
\Xi=&\sum_{N_+=0}^\infty\sum_{N_-=0}^\infty
\frac{z_+^{N_+}z_-^{N_-}}{N_+!N_-!}\idotsint\limits_{V^{N_++N_-}}
\ud\mb{r}^{N_++N_-}\\&\times{}\mathrm{exp}\lbrace-\frac{\beta}{2}
\sum_{\substack{i,j=1\\i\ne{}j}}^{N_++N_-}\lbrack\frac{\sigma_i
\sigma_j}{r_{i,j}}+V_{hc}(r_{i,j},\sigma_i,
\sigma_j|a,\lambda)\rbrack\rbrace
\end{split}
\end{equation}
where $\sigma_i$ is the sign of ion $i$ and
$r_{i,j}=\norm{\mathbf{r}_i-\mathbf{r}_j}$ is the distance between
ion $i$ and ion $j$. The hard core interaction $V_{hc}$ has the
usual meaning, here we have to specify the signs of the two ions
involved because of the size asymmetry.

We now carry out the CP procedure on any configuration of the
system, and group pairs that have the inter-ion distance in
$[a,a+\xi]$, $\xi$ being infinitesimal; denote the number of such
pairs by $N_0$. For the rest of the ions in the system, this
operation and DNR would change the first two system parameters
from $(a,\lambda)$ to $(a+\xi,\tilde{\lambda})$, where
$\tilde{\lambda}=\mathrm{min}(1, \lambda+\xi(1+\lambda)/a)$. Now
we have:
\begin{equation}\label{PartitionScaleUp}
\begin{split}
\Xi\doteq{}&\sum_{N_+=0}^\infty\sum_{N_-=0}^\infty
\frac{z_+^{N_+}z_-^{N_-}}{N_+!N_-!}\idotsint\limits_{V^{N_++N_-}}
\ud\mb{r}^{N_++N_-}\\&\times{}\mathrm{exp}\lbrace-\frac{\beta}{2}
\sum_{\substack{i,j=1\\i\ne{}j}}^{N_++N_-}\lbrack\frac{\sigma_i
\sigma_j}{r_{i,j}}+V_{h.c.}(r_{i,j},\sigma_i,\sigma_j|a+\xi,
\tilde{\lambda})\rbrack\rbrace\\
&\times\sum_{N_0=0}^\infty\frac{(z_+z_-)^{N_0}}{N_0!}\idotsint
\limits_{(V\times{}v)^{N_0}}\ud\mathbf{X}^{N_0}
\mathrm{exp}(\sum_{k=1}^{N_0}\frac{\beta}{s_k})\\
&\times{}\mathrm{exp}\lbrace-\beta{}\sum_{l=1}^{N_0}\lbrack
\sum_{m=1}^{N_++N_-}V_{ps,\,l,\,m}
+\sum_{\substack{n=1\\n\ne{}l}}^{N_0}\frac{V_{pp,\,l,\,n}}
{2}\rbrack\rbrace
\end{split}
\end{equation}
where $\mathbf{X}=\mathbf{R}\oplus{}\mathbf{s}$ is the
six-component vector comprising the position of the center of a
pair, $\mathbf{R}$, and the relative displacement vector,
$\mathbf{s}$, which points from $-$ ion to $+$ ion of this pair.
In the integration range of $\mathbf{X}$, $v$ is the range of
$\mathbf{s}$ delimited by the requirement that $s\in[a,a+\xi]$.
$V_{ps,\,l,\,m}$ denotes the interaction between pair $l$ and
single ion $m$, written explicitly:
\begin{displaymath}\label{Vps}
\begin{split}
V_{ps,\,l,\,m}\equiv&V_{ps}(\mathbf{X}_l,\mb{r}_m,\sigma_m|a,\lambda)
\equiv{}V_{ps}(\mathbf{R}_l,\mathbf{s}_l,\mb{r}_m,\sigma_m|a,\lambda)\\
=&V_{hc}(\norm{(\mb{R}_l+\frac{\mb{s}_l}{2})-\mb{r}_m},+1,
\sigma_m|a,\lambda))\\
&+V_{hc}(\norm{(\mb{R}_l-\frac{\mb{s}_l}{2})-\mb{r}_m},-1,
\sigma_m|a,\lambda))\\
&+V_{cd}(\mathbf{R}_l,\mathbf{s}_l,\mb{r}_m,\sigma_m);
\end{split}
\end{displaymath}
where $V_{cd}$ is the extra spacial exclusion imposed by CP and
DNR:
\begin{displaymath}
V_{cd}(\mathbf{R}_l,\mathbf{s}_l,\mb{r}_m,\sigma_m)=\left
\{\begin{array}{ll} \infty & \textrm{if }
\norm{\mb{R}\pm\sigma_m(\frac{\mb{s}_l}{2})-\mb{r}_m}<s_l;
\\ 0 & \textrm{otherwise}.
\end{array}\right.
\end{displaymath}
Notice that in the expression of $V_{ps}$ there are no electric
interaction terms because of CPA, and this makes $V_{ps}$ a
short-ranged interaction. $V_{pp,\,l,\,n}$ is the interactions
between two CP pairs, $l$ and $n$, which is also short-ranged
because of CPA. Since $\xi$ is infinitesimal, configurations with
more than one chosen pair in any finite region can be neglected,
so can $V_{pp}$, and the summation over $N_0$ in
Equation~\eqref{PartitionScaleUp} can be replaced by an
exponential form:
\begin{equation}\label{PartitionScaleUpExp}
\mathrm{exp}\lbrace\xi{}z_+z_-e^{\frac{\beta}{a}}a^2\times
\mathcal{M}\rbrace
\end{equation}
where
\begin{equation}\label{M}
\mathcal{M}\!=\!\int\limits_V\!\ud{}\mb{R}\int\limits_{4\pi}
\!\ud\omega\,
\mathrm{exp}\lbrace-\beta{}\!\sum_{m=1}^{N_++N_-}\!V_{ps}
(\mathbf{R},\mathbf{a}(\omega),\mb{r}_m,\sigma_m|a,\lambda)\rbrace
\end{equation}
$\omega$ is the orientation of the pair supplying the direction of
$\mb{a}(\omega)$ whose magnitude is $a$.

To evaluate Equation~\eqref{M}, we decompose the exponential form
in the usual way:
\begin{equation}\label{Mdecompse}
\begin{split}
\mathcal{M}\!&=\!\int\limits_V\!\ud{}\mb{R}\int\limits_{4\pi}
\!\ud\omega\,\prod_{m=1}^{N_++N_-}(f_m+1)\\
&=\mathcal{N}+\sum_{m=1}^{N_++N_-}\mathcal{S}_{m}
+\sum_{\substack{n,n'=1\\n\ne{}n'}}^{N_++N_-}\mathcal{D}_
{n,n'}+\ldots
\end{split}
\end{equation}
where:
\begin{displaymath}
\begin{split}
f_m&=\mathrm{exp}\lbrace-\beta{}V_{ps}(\mathbf{R},\mathbf{a}
(\omega),\mb{r}_m,\sigma_m|a,\lambda)\rbrace-1;\\
\mathcal{N}&=\!\int\limits_V\!\ud{}\mb{R}\int\limits_{4\pi}
\!\ud\omega\,1=4\pi{}V;\\
\mathcal{S}_m&=\!\int\limits_V\!\ud{}\mb{R}\int\limits_{4\pi}
\!\ud\omega{}f_m;\\
\mathcal{D}_{n,n'}&=\!\int\limits_V\!\ud{}\mb{R}\int\limits_
{4\pi}\!\ud\omega(f_n\times{}f_{n'}).
\end{split}
\end{displaymath}
Higher order terms which involve more than two $f$'s can be
defined similarly. Notice that $\mathcal{S}$ is the same for all
ions of the same sign, in the following we'll use
$\mathcal{S}_+(a,\lambda)$ and $\mathcal{S}_-(a,\lambda)$ to
denote them respectively. $\mathcal{D}_{n,n'}$ depends on the
value of $(a,\lambda)$ and the relative position and signs of the
ion $n$ and $n'$; it acts like an extra 2-body interaction for the
rest of the ions in the system. Similarly higher order terms act
like extra multi-body interactions. Now the partition function
becomes:
\begin{equation}\label{Partition1}
\begin{split}
\Xi\doteq{}&\mathrm{exp}\lbrace\xi{}z_+z_-e^{\frac{\beta}{a}}a^2
\times\mathcal{N}\rbrace\\
&\times\sum_{N_+=0}^\infty\sum_{N_-=0}^\infty
\frac{\tilde{z}_+^{N_+}\tilde{z}_-^{N_-}}{N_+!N_-!}\idotsint
\limits_{V^{N_++N_-}}\ud\mb{r}^{N_++N_-}\\
&\times{}\mathrm{exp}\lbrace-\frac{\beta}{2}\sum_{\substack
{i,j=1\\i\ne{}j}}^{N_++N_-}\lbrack\frac{\sigma_i\sigma_j}{r_{i,j}}
+V_{h.c.}(\sigma_i,\sigma_j,r_{i,j}|a+\xi,\tilde{\lambda})
\rbrack\rbrace\\&\times{}\mathrm{exp}\lbrace\xi{}z_+z_-e^
{\frac{\beta}{a}}a^2\times\sum_{\substack{n,n'=1\\n\ne{}n'}}
^{N_++N_-}\mathcal{D}_{n,n'}+\ldots\rbrace
\end{split}
\end{equation}
where
$\tilde{z}_{\pm}=z_\pm\times{}\mathrm{exp}\lbrace\xi{}z_+z_-e^
{\frac{\beta}{a}}a^2\times\mathcal{S}_\pm(a,\lambda)\rbrace$.

We see an interesting pattern in Equation~\eqref{Partition1}: the
LHS denotes the partition function with system parameter
$(a,\lambda,z_+,z_-,\beta)$, while the RHS has an exponential term
followed by the expression of a partition function with system
parameter $(a+\xi,\tilde{\lambda},\tilde{z}_+,\tilde{z}_-,\beta)$
and the extra interactions introduced by $\mc{D}$ and high order
terms of $\mc{M}$. We can of course assume that there are also
these extra interactions on the LHS, but they have the special
initial value of 0. Because of this similarity, the same procedure
we performed on the original partition function can be carried out
on the new partition function on the RHS and then repeated: a
differential procedure that resembles the Renormalization Group
Theory, except that no fixed point of the system parameters is
expected. Since $V_{ps}$ is short-ranged, we will neglect $\mc{D}$
and all the higher order terms of $\mc{M}$ in the following for
simplicity; their effects will be discussed later in this report.
The system parameters are now interpreted as variables to be
"scaled up" in this procedure, with no confusion, denoted as
$(\tilde{a},\tilde{\lambda},\tilde{z}_+,\tilde{z}_-,\tilde{\beta})$
that take the initial values of $(a,\lambda,z_+,z_-,\beta)$, and
$\tilde{a}$ serves as the scaling factor. From the above
derivation, the differential equations of the other system
parameters with respect to $\tilde{a}$ are:
\begin{equation}\label{DifferentialEquation}
\begin{split}
\tilde{\lambda}'(\tilde{a})&=\left\{\begin{array}{cl}
(1+\tilde{\lambda})/\tilde{a} \quad &\textrm{if}\,
\tilde{\lambda}<1;
\\ 0& \textrm{otherwise};
\end{array}\right.\\
\tilde{z}'_\pm{}(\tilde{a})&=\tilde{z}_\pm\times{}\tilde{z}
_+\tilde{z}_-e^{\frac{\beta}{\tilde{a}}}\tilde{a}^2\times
\mathbf{S}_\pm(\tilde{a},\tilde{\lambda});\\
\tilde{\beta}'(\tilde{a})&=0.
\end{split}
\end{equation}
which are easily solved. This procedure can be carried out until
the temperature is not considered low anymore when
$\tilde{a}=\Lambda\sim{}\beta$, and we have:
\begin{equation}\label{PartitionIntegration}
\begin{split}
\Xi\doteq{}&\mathrm{exp}\lbrace{}V\smallint_{a}^\Lambda{}
\tilde{z}_+(\tilde{a})\tilde{z}_-(\tilde{a})e^{\frac{\beta}
{\tilde{a}}}4\pi{}\tilde{a}^2\ud{}\tilde{a}\rbrace\times\\
&\Xi(\Lambda,\tilde{\lambda}(\Lambda),\tilde{z}_+(\Lambda),
\tilde{z}_-(\Lambda),\beta).
\end{split}
\end{equation}
where $\Xi(\Lambda,\tilde{\lambda}(\Lambda),\tilde{z}_+(\Lambda),
\tilde{z}_-(\Lambda),\beta)$ is the contribution from "free
moving" ions, which can be evaluated by linearized theories such
as the Mean Spherical Approximation (MSA) or the Debye-Huckle (DH)
Theory. The exact value of $\Lambda$ is, of course, not important;
in this report, we set $\Lambda=\beta/2$.

Equation \eqref{PartitionIntegration} concludes the above
derivation, with simplifications of course: recall that
$\mathcal{D}$ and higher order terms of $\mathcal{M}$ have been
neglected. If these terms are included in the derivation,
theoretically speaking, a similar differential procedure can still
be constructed, but it will have greater complication. Such
simplification overestimates the repulsion between the CP pair and
single ions at each step of the derivation, consequently, the
value of pressure is raised; and the contribution from
$\Xi(\Lambda,\tilde{\lambda}(\Lambda),\tilde{z}_+(\Lambda),
\tilde{z}_-(\Lambda),\beta)$ is lower estimated because
$\tilde{z}_{\pm}(\tilde{a})$ is decreasing faster than it ought to
be. But if the density of the system is not too high, it should
not significantly affect the result. And if we are interested in
situations with very low temperature: $\beta\gg\rho^{-1/3}$, how
it is when the system is close to the critical point of the phase
transition, the contribution from "free moving" ions can be
neglected, because, not surprisingly, almost all ions would be
closely paired up at such a low temperature; this can also be seen
from the fact that, when $\beta\gg{}a$ and $z_{\pm}$ is
sufficiently big (so that $\rho^{-1/3}\ll\beta$),  if
$\Xi(\Lambda,\tilde{\lambda}(\Lambda),\tilde{z}_+(\Lambda),
\tilde{z}_-(\Lambda),\beta)$ is evaluated in the same fashion as
above (the integration with respect to $\tilde{a}$ now goes from
$\Lambda$ to $\infty$), with CPA but without DNR: a clear
overestimation, its contribution is still negligible
\cite{MyThesis}. The partition function then takes a very simple
form, written explicitly:
\begin{equation}\label{PartitionFinal}
\Xi\doteq{}\mathrm{exp}\lbrace{}V\smallint_{a}^\Lambda\frac
{ze^{\frac{\beta}{\tilde{a}}}4\pi{}\tilde{a}^2}
{1-z\smallint_a^{\tilde{a}}g(x)\ud{}x} \ud{}\tilde{a}\rbrace
\end{equation}
where $z=z_+z_-$, and:
\begin{equation}\label{g}
g(x)=x^2e^{\beta/x}[\mc{S}_+(x,\tilde{\lambda}(x))+\mc{S}_-(x,
\tilde{\lambda}(x))]
\end{equation}

The description of the system by Equation \eqref{PartitionFinal},
however, does not exhibit a phase transition. Physically, we have
seen that when CP pairs are evenly (randomly) distributed,
$I\approx{}I_{cp}$ and the electric interaction between CP pairs,
denoted by $V_{ppe}$ in the following, overall doesn't have
significant effects due to cancellation. But this ceases to be
true for CP pairs in clusters, which are shown by MC simulations
(e.g. \cite{Enrique, SAPM}) to form abundantly when the system is
close to phase transition. Recall the well-known relationship:
$U=\frac{1}{8\pi}\int{}\mathbf{E}^2(\mb{r})\mathrm{d}\mb{r}$, for
two CP pairs, lowering $V_{ppe}$ would lower $U$, which in turn
lowers their electric field as a whole and also makes the "shape"
of their $\mathbf{E}(\mb{r})$ more concentrated. Because their
electric field serves as the agent of the interaction between the
two CP pairs and the rest of the system, lowering $V_{ppe}$ would
then make them more isolated from the rest of the system and their
negative interaction less likely to be cancelled by positive
interactions. For larger clusters, this effect becomes even more
significant. In the following, we construct a VdW-like term to
estimate the effect of this "extra" attraction between CP pairs
and extend the scheme to exhibit phase transition.

The previous derivation is treated as the leading order
calculation, from which the mean inter-ion separation of CP pairs
is found to be \cite{MyThesis}:
\begin{equation}\label{save}
\begin{split}
\langle{}s\rangle=\frac{2}{\rho}\int_{a}^\Lambda\!\frac
{ze^{\frac{\beta}{\tilde{a}}}
[1-z\smallint_a^{\tilde{a}}g(y)(1-\frac{y}{\tilde{a}})
\ud{}y]4\pi{}\tilde{a}^3}
{[1-z\smallint_a^{\tilde{a}}g(x)\ud{}x]^2}\,\ud{}\tilde{a}
\end{split}
\end{equation}
The change of pressure as the result of including this "extra"
pair attraction is then estimated to be:
\begin{equation}\label{P2viral}
\Delta{}P_2=\frac{1}{\beta}B_{2,\,e}(\langle{}s\rangle,\lambda,
\beta)(\frac{\rho}{2})^2
\end{equation}
where
\begin{equation}\label{B2}
\begin{split}
B_{2,\,e}(\langle{}s\rangle,\lambda,\beta)=&-\frac{1}{2V}\!\int
\limits_V\!\ud{}\mathbf{R}_1\!\int\limits_{4\pi}\!\frac{\ud\mathbf
{\omega}_1}{4\pi}\times\\&\iint\limits_{\substack{V_{ppe}<\mathcal
{E}_2\\}}\!\frac{\ud{}\mathbf{R}_2\ud\mb{\omega}_2}{4\pi}
\Big\lbrace{}e^{-\beta{}V_{ppe}}-1\Big\rbrace
\end{split}
\end{equation}
where
$V_{ppe}\!\!\equiv\!\!{}V_{ppe}(\mathbf{R}_1,\mb{\omega},\mathbf{R}_2,
\mb{\omega}_2|\langle{}s\rangle,\lambda)$ gives the electric
interaction between two identical CP pairs of inter-ion separation
$\langle{}s\rangle$ for configurations allowed by CP and DNR, and
otherwise 0. Here, $B_{2,\,e}$ accounts for the "extra" attraction
introduced by 4-ion (two CP pairs) clusters, and, following the
spirit of the above discussion, we propose the use of energy
standard in defining clusters: $\mathcal{E}_2$ is the maximum
electric interaction between two pairs for them to be considered
as a 4-ion cluster, bigger clusters can be defined in a similar
fashion.The values of $\mathcal{E}_2$ and $\mathcal{E}$s for
clusters involving more than two CP pairs are chosen so that, in
the sense of statistical average, when $I<(-1/a)$ (recall
$I_{cp}\geqslant(-1/a)$), $(I_{cp}+I_{cluster})$ "closely"
approximate $I$, where $I_{cluster}$ denotes the "extra"
attraction introduced by clusters. $\mathcal{E}$s would of course
depend on the temperature and the density of the system. Here, we
take a short cut and choose
$\mathcal{E}_2\!=\!-0.48/\langle{}s\rangle$ so that the critical
temperature of RPM matches the result of MC simulations, which has
converged to $\sim{}0.050(q_0^2/k_BDa)$ in recent years.

Since the big asymmetry cases raise new problems with the
formation of special micro-structures \cite{SAPM} and is still
under investigation, here we focus on situations with only
moderate asymmetry. Some of the critical points obtained from this
work are reported in Table~\ref{CriticalPoint}, in which
$T_c^*=k_BT_cDa/q_0^2$ and $\rho_c^*=\rho_ca^3$ are the reduced
critical temperature and reduced critical density respectively,
values from the MC simulation of \cite{SAPM} , if available, are
also listed in the table as comparison, they are denoted by
$T_{c,Y}^*$ and $\rho_{c,Y}^*$.
\begin{table}[h]
\begin{center}
\caption{locations of critical points\label{CriticalPoint}}
\begin{tabular*}{0.45\textwidth}{@{\extracolsep{\fill}}ccccc}
\hline \hline
$\lambda$ & $T_c^*$ & $\rho_c^*$ & $T_{c,Y}^*$ & $\rho_{c,Y}^*$\\
\hline
1.00&0.049(6)&0.070(3)&0.0492(2)&0.073(2)\\
0.75&0.049(2)&0.063(4)&0.0488(2)&0.072(2)\\
0.50&0.048(2)&0.053(5)&0.0475(3)&0.070(2)\\
0.45&0.047(7)&0.051(2)&-&-\\
0.40&0.047(0)&0.048(6)&-&-\\
0.35&0.045(9)&0.046(1)&-&-\\
0.30&0.044(3)&0.043(5)&-&-\\
0.25&0.041(6)&0.040(6)&0.0422(3)&0.059(3)\\
0.20&0.036(7)&0.038(0)&0.0386(4)&0.051(3)\\
\hline \hline
\end{tabular*}
\end{center}
\end{table}

For the critical density of RPM, recent MC simulations have
converged to $\rho_c^*\sim0.075$, the result of this work (the
case with $\lambda=1.00$) has a discrepancy of less than $10\%$.
For the critical temperatures of all the values of $\lambda$, we
see a very good agreement (less than $10\%$ of discrepancy) with
the simulation results of \cite{SAPM}. More importantly, the
result shows that both critical temperature and critical density
decrease as the size asymmetry increases, and the decrease is
small until $\lambda\sim{}0.4$, after which they decrease more
dramatically, these features are consistent with the results of MC
simulations\cite{SAPM, Enrique}; to our knowledge, it has not been
shown before theoretically.

Jerome K. Percus acknowledges the support from the Department of
Energy (DOE).

\end{document}